\documentclass[twocolumn,showpacs,aps,prl,superscriptaddress,preprintnumbers,amsmath,amssymb,floatfix]{revtex4}

\usepackage{graphicx}    
\usepackage{dcolumn}     
\usepackage{bm}          

\usepackage{psfrag}

\newcommand{\BABARPubYear}    {05}
\newcommand{\BABARPubNumber}  {004}

\newcommand{\SLACPubNumber}   {11085}

\input babarsym

\long\def\inst#1{\par\nobreak\kern 4pt\nobreak
    {\it #1}\par\vskip 10pt plus 3pt minus 3pt}

\RequirePackage{xspace}

\usepackage{relsize}
\def\babar{\mbox{\slshape B\kern-0.1em{\smaller A}\kern-0.1em
    B\kern-0.1em{\smaller A\kern-0.2em R}}}

\def\en         {\ensuremath{e^-}\xspace}   
\def\ep         {\ensuremath{e^+}\xspace}
 
\def\epem       {\ensuremath{e^+e^-}\xspace}

\def\mup        {\ensuremath{\mu^+}\xspace}
\def\mun        {\ensuremath{\mu^-}\xspace} 
\def\mumu       {\ensuremath{\mu^+\mu^-}\xspace}

\def\taup       {\ensuremath{\tau^+}\xspace}
\def\taum       {\ensuremath{\tau^-}\xspace}

\def\ellell     {\ensuremath{\ell^+ \ell^-}\xspace}
\def\nub        {\ensuremath{\overline{\nu}}\xspace}
\def\nunub      {\ensuremath{\nu{\overline{\nu}}}\xspace}
\def\nub        {\ensuremath{\overline{\nu}}\xspace}
\def\nunub      {\ensuremath{\nu{\overline{\nu}}}\xspace}
\def\nue        {\ensuremath{\nu_e}\xspace}
\def\nueb       {\ensuremath{\nub_e}\xspace}

\def\num        {\ensuremath{\nu_\mu}\xspace}
\def\numb       {\ensuremath{\nub_\mu}\xspace}

\def\nut        {\ensuremath{\nu_\tau}\xspace}
\def\nutb       {\ensuremath{\nub_\tau}\xspace}

\def\nul        {\ensuremath{\nu_\ell}\xspace}
\def\nulb       {\ensuremath{\nub_\ell}\xspace}

\def\g     {\ensuremath{\gamma}\xspace}
\def\gaga  {\ensuremath{\gamma\gamma}\xspace}  
\def\ggstar{\ensuremath{\gamma\gamma^*}\xspace}

\def\qqbar {\ensuremath{q\overline q}\xspace}

\def\piz   {\ensuremath{\pi^0}\xspace}

\def\ppz   {\ensuremath{\pi^0\pi^0}\xspace}
\def\pip   {\ensuremath{\pi^+}\xspace}
\def\pim   {\ensuremath{\pi^-}\xspace}
\def\pipi  {\ensuremath{\pi^+\pi^-}\xspace}

\def\pimp  {\ensuremath{\pi^\mp}\xspace}
\def\kaon  {\ensuremath{K}\xspace}
\def\Kbar  {\kern 0.2em\overline{\kern -0.2em K}{}\xspace}

\def\Kz    {\ensuremath{K^0}\xspace}
\def\Kzb   {\ensuremath{\Kbar^0}\xspace}
\def\KzKzb {\ensuremath{\Kz \kern -0.16em \Kzb}\xspace}
\def\Kp    {\ensuremath{K^+}\xspace}
\def\Km    {\ensuremath{K^-}\xspace}
\def\Kpm   {\ensuremath{K^\pm}\xspace}

\def\KpKm  {\ensuremath{\Kp \kern -0.16em \Km}\xspace}
\def\KS    {\ensuremath{K^0_{\scriptscriptstyle S}}\xspace} 
\def\KL    {\ensuremath{K^0_{\scriptscriptstyle L}}\xspace} 
\def\Kstarz  {\ensuremath{K^{*0}}\xspace}

\def\Kstar   {\ensuremath{K^*}\xspace}

\def\Kstarp  {\ensuremath{K^{*+}}\xspace}

\def\Dbar    {\kern 0.2em\overline{\kern -0.2em D}{}\xspace}

\def\Dz      {\ensuremath{D^0}\xspace}
\def\Dzb     {\ensuremath{\Dbar^0}\xspace}
\def\DzDzb   {\ensuremath{\Dz {\kern -0.16em \Dzb}}\xspace}
\def\Dp      {\ensuremath{D^+}\xspace}
\def\Dm      {\ensuremath{D^-}\xspace}

\def\DpDm    {\ensuremath{\Dp {\kern -0.16em \Dm}}\xspace}
\def\Dstar   {\ensuremath{D^*}\xspace}

\def\Dstarz  {\ensuremath{D^{*0}}\xspace}

\def\Dstarp  {\ensuremath{D^{*+}}\xspace}
\def\Dstarm  {\ensuremath{D^{*-}}\xspace}

\def\B       {\ensuremath{B}\xspace}
\def\Bbar    {\kern 0.18em\overline{\kern -0.18em B}{}\xspace}
\def\Bb      {\ensuremath{\Bbar}\xspace}
\def\BB      {\ensuremath{\B {\kern -0.16em \Bb}}\xspace}
\def\Bz      {\ensuremath{B^0}\xspace}
\def\Bzb     {\ensuremath{\Bbar^0}\xspace}
\def\BzBzb   {\ensuremath{\Bz {\kern -0.16em \Bzb}}\xspace}
\def\NBB     {\ensuremath{N_{\BB}}}
\def\Bu      {\ensuremath{B^+}\xspace}
\def\Bub     {\ensuremath{B^-}\xspace}

\def\Bpm     {\ensuremath{B^\pm}\xspace}

\def\BpBm    {\ensuremath{\Bu {\kern -0.16em \Bub}}\xspace}

\def\BorBbar    {\kern 0.18em\optbar{\kern -0.18em B}{}\xspace}
\def\DorDbar    {\kern 0.18em\optbar{\kern -0.18em D}{}\xspace}
\def\KorKbar    {\kern 0.18em\optbar{\kern -0.18em K}{}\xspace}

\def\jpsi     {\ensuremath{{J\mskip -3mu/\mskip -2mu\psi\mskip 2mu}}\xspace}

\mathchardef\Upsilon="7107
\def\Y#1S{\ensuremath{\Upsilon{(#1S)}}\xspace}

\def\FourS {\Y4S}

\mathchardef\Deltares="7101
\mathchardef\Xi="7104
\mathchardef\Lambda="7103
\mathchardef\Sigma="7106
\mathchardef\Omega="710A

\def\Deltabar{\kern 0.25em\overline{\kern -0.25em \Deltares}{}\xspace}
\def\Lbar{\kern 0.2em\overline{\kern -0.2em\Lambda\kern 0.05em}\kern-0.05em{}\xspace}
\def\Sigbar{\kern 0.2em\overline{\kern -0.2em \Sigma}{}\xspace}
\def\Xibar{\kern 0.2em\overline{\kern -0.2em \Xi}{}\xspace}
\def\Obar{\kern 0.2em\overline{\kern -0.2em \Omega}{}\xspace}
\def\Nbar{\kern 0.2em\overline{\kern -0.2em N}{}\xspace}
\def\Xb{\kern 0.2em\overline{\kern -0.2em X}{}\xspace}
\def\X {\ensuremath{X}\xspace}

\def\bpsikl     {\ensuremath{\Bz \to \jpsi \KL}\xspace}

\def\Bzbtox     {\ensuremath{\Bzb \to \X}\xspace}

\def\invfb   {\ensuremath{\mbox{\,fb}^{-1}}\xspace}

\def\mus  {\ensuremath{\rm \,\mus}\xspace}

\def\mus        {\ensuremath{\,\mu{\rm s}}\xspace}    

\def\degrees{\ensuremath{^{\circ}}\xspace}

\def\to                 {\ensuremath{\rightarrow}\xspace}

\def\pep2{PEP-II}
\def\BF{$B$ Factory}

\def\gsim{{~\raise.15em\hbox{$>$}\kern-.85em
          \lower.35em\hbox{$\sim$}~}\xspace}
\def\lsim{{~\raise.15em\hbox{$<$}\kern-.85em
          \lower.35em\hbox{$\sim$}~}\xspace}

\def\epstag  {\ensuremath{\varepsilon_{\rm tag}}\xspace}

\def\jetset74   {\mbox{\tt Jetset \hspace{-0.5em}7.\hspace{-0.2em}4}\xspace}

\usepackage{relsize}

\def\Pv{\ensuremath{{\cal P}_V}}
\def\Mnu{\ensuremath{{\cal M}^2}}

\def\Ms{\ensuremath{{\cal M}_s^2}}
\def\M{\ensuremath{{\cal M}_1^2}}
\def\MM{\ensuremath{{\cal M}_2^2}}

\def\fzz{\ensuremath{f_{00}}}
\def\Ns{\ensuremath{N_s}}
\def\Nd{\ensuremath{N_d}}

\begin{document}

\begin{flushleft}
{\babar-PUB-\BABARPubYear/\BABARPubNumber \\
SLAC-PUB-\SLACPubNumber      \\
\hfill April 2005}           \\
\end{flushleft}

\title{Measurement of the Branching Fraction of
{\boldmath{$\Upsilon(4S) \rightarrow \BzBzb$}}}

%
\author{B.~Aubert}
\author{R.~Barate}
\author{D.~Boutigny}
\author{F.~Couderc}
\author{Y.~Karyotakis}
\author{J.~P.~Lees}
\author{V.~Poireau}
\author{V.~Tisserand}
\author{A.~Zghiche}
\affiliation{Laboratoire de Physique des Particules, F-74941 Annecy-le-Vieux, France }
\author{E.~Grauges-Pous}
\affiliation{IFAE, Universitat Autonoma de Barcelona, E-08193 Bellaterra, Barcelona, Spain }
% ...
\collaboration{The \babar\ Collaboration}
\noaffiliation

\date{\today} 

\begin{abstract}
We report the first measurement of the branching fraction 
\fzz\ for $\Upsilon(4S) \rightarrow \BzBzb$.  The data sample consists of 81.7\invfb collected 
at the $\Y4S$ resonance with the \babar\ detector at the \pep2\ asymmetric-energy 
$e^+e^-$ storage ring.  Using partial reconstruction of the decay $\Bzb \rightarrow D^{*+}
\ell^{-} \bar{\nu}_{\ell}$ in which only the charged lepton and the soft pion from the decay
$D^{*+}\to \Dz \pi^+$ are reconstructed, we obtain 
$\fzz = 0.487 \pm 0.010(stat) \pm 0.008(sys)$.
Our result does not depend on the branching fractions of 
$\Bzb \rightarrow D^{*+} \ell^{-} \bar{\nu}_{\ell}$ and $D^{*+} \to \Dz \pi^+$ decays, 
on the ratio of the charged and neutral $B$ meson lifetimes, nor on the assumption 
of isospin symmetry.
\end{abstract}

\maketitle

Isospin violation in the decay $\Upsilon(4S) \rightarrow \BB$ will lead to a difference
between the branching fractions $\fzz \equiv {\cal B}(\Upsilon(4S) \rightarrow \BzBzb)$ and
$f_{+-} \equiv {\cal B}(\Upsilon(4S) \rightarrow B^+ {B}^{-})$.
Predictions for the ratio $R^{+/0} \equiv f_{+-} / \fzz$ range from 1.03 to 1.25~\cite{eichten}.
Measurements of $R^{+/0}$~\cite{haleh,babar,belle03,godang02,silvia} have
been made assuming isospin symmetry in specific decay rates and resulting in an average value of
$1.006 \pm 0.039$~\cite{hfag05}, consistent with isospin conservation in $\Upsilon(4S)$ 
decays to $\BB$.  To date no measurement has been made of either $\fzz$ or $f_{+-}$.  
In this paper we report the first direct measurement of $\fzz$. 
It is completely independent of the previous measurements of $R^{+/0}$.
Independent measurements of $\fzz$ and $R^{+/0}$ can be used to constrain
the $\Upsilon(4S) \rightarrow$ non-$\BB$ fraction.
The $\fzz$ value is important for measuring absolute $\Upsilon(4S)$ branching
fractions and for measuring $V_{cb}$, the Cabibbo-Kobayashi-Maskawa matrix element.

The data sample used in this analysis consists of 81.7\invfb collected 
at the $\Y4S$ resonance (on-resonance) and 9.6\invfb collected 40\mev 
below the resonance (off-resonance).  
The on-resonance data sample has a mean energy of 10.580\gev and an energy rms spread of 
4.6\mev. Due to the small spread, any plausible energy
dependence of \fzz\ has a negligible effect on the central value.
A simulated sample of $\BB$ with integrated luminosity equivalent to approximately three 
times the data is used for background studies.

A detailed description of the \babar\ detector and the algorithms used
for track reconstruction and particle identification is provided
elsewhere~\cite{babar_nim}. A brief summary is given here.
High-momentum particles are reconstructed by
matching hits in the silicon vertex tracker (SVT) with track elements
in the drift chamber (DCH). Lower momentum tracks, which do not leave
signals on many wires in the DCH due to the bending induced by a
magnetic field, are reconstructed in the SVT alone.
Electrons are identified by the ratio of the track momentum to the
associated energy deposited in the calorimeter (EMC), the transverse
profile of the shower, the energy loss in the drift chamber, and
information from a Cherenkov detector (DIRC).  
Muons are identified in the instrumented flux return (IFR), composed
of resistive plate chambers and layers of iron.
Muon candidates are required to have a path length and hit
distribution in the IFR and energy deposition in the EMC consistent 
with that expected for a minimum-ionizing particle.
The \babar\ detector Monte Carlo simulation is based on GEANT4~\cite{geant4}.

We select the decays $\Bzb \rightarrow D^{*+} \ell^{-} \bar{\nu}_{\ell}$, 
$D^{*+} \to \Dz \pi^+$ ($\ell = e,\,\mu$).  
The inclusion of charge-conjugate reactions is implied throughout this paper. 
The sample of events in which at least one $\Bzb \rightarrow D^{*+} \ell^{-}
\bar{\nu}_{\ell}$ candidate decay is found is labeled the ``single-tag sample''. 
The number of signal decays in this sample is
\begin{equation}
\Ns =
      2 \NBB \fzz\, \varepsilon_{s} \, 
      {\cal B}(\Bzb \rightarrow D^{*+} \ell^- \bar{\nu}_{\ell}),
\label{eq:ns}
\end{equation} 
where $\NBB$ is the total number of $\BB$ events in the data sample and $\varepsilon_{s}$ 
is the reconstruction efficiency for $\Bzb \rightarrow D^{*+} \ell^{-} \bar{\nu}_{\ell}$.  
We determine $\NBB$ = 88.7 million events by counting the number of hadronic decays in 
the on-resonance data and subtracting the $\epem\to\qqbar$ ($q$ = $u$, $d$, $s$, or $c$ quark) component 
using off-resonance data, as described in detail in Ref.~\cite{bcounting}.  The error in $\NBB$ is $1.1\%$ and 
is dominated by systematic uncertainties.  We attribute all \BB pairs to $\Upsilon(4S)$ decays.

The number of signal events in the subset in which two $\Bzb \rightarrow D^{*+} 
\ell^{-}\bar{\nu}_{\ell}$ candidates are found is labeled the ``double-tag sample''. 
The number of such events is
\begin{equation}
\Nd =  
      \NBB \, \fzz\, \varepsilon_{d} \,
      [{\cal B}(\Bzb \rightarrow D^{*+} \ell^- \bar{\nu}_{\ell})]^2 ,  
\label{eq:nd}
\end{equation} 
where $\varepsilon_{d}$ is the efficiency to reconstruct two $\Bzb \rightarrow D^{*+}
\ell^{-}\bar{\nu}_{\ell}$ decays in the same event.
From Eq.~(\ref{eq:ns}) and Eq.~(\ref{eq:nd}), \fzz\ is given by
\begin{equation}
\fzz =  
       {C N_{s}^{2} \over {4 N_{d} \NBB} },
\label{eq:f00}
\end{equation}
where we have defined $C \equiv \varepsilon_{d} / \varepsilon_{s}^{2}$. 
The value of $C$ is 1 if the efficiencies for detecting each $B$ meson are 
uncorrelated in double-tag events, which, given the pseudoscalar nature of $B$
mesons and the proximity of the $\Upsilon(4S)$ to the $\BB$ threshold, is expected.
Using the Monte Carlo simulation we determine $C = 0.995 \pm 0.008$, 
where the error is due to the finite size of the simulated sample.

We select the decays $\Bzb \rightarrow D^{*+} \ell^{-} \bar{\nu}_{\ell}$ 
with a partial reconstruction technique~\cite{godang02,argus,delphi,franco}.  
In this technique, only the lepton from the decay $\Bzb \rightarrow D^{*+} 
\ell^{-} \bar{\nu}_{\ell}$ and the soft pion from the decay
$D^{*+}\to \Dz \pi^+$ are reconstructed. No attempt is made to reconstruct
the $\Dz$, resulting in a high reconstruction efficiency. 

The $\Bzb$ decay point is determined from a vertex fit of the soft-pion
and lepton tracks, with the vertex constrained to the beam spot position in 
the $x-y$ plane. We only use events with vertex-fit probability, \Pv, greater 
than $0.1\%$ to optimize a signal-to-background ratio. 

We select hadronic events by requiring at least four charged particle tracks 
reconstructed in the SVT and the DCH. 
To reduce non-$\BB$ background, the ratio of the second to the zeroth Fox-Wolfram 
moments~\cite{wolfram}, $R_2 = H_2/H_0$, is required to be less than 0.5.

To suppress leptons from charm decays, all lepton candidates are required to have 
momenta between 1.5\gevc and 2.5\gevc in the $\epem$ center-of-mass frame.
Soft pion candidates are required to have center-of-mass momenta
between 60\mevc and 200\mevc.
As a consequence of the limited phase space available in the $D^{*+}$
decay, the soft pion is emitted nearly at rest in the $D^{*+}$ rest frame.
The $D^{*+}$ four-momentum can therefore be computed by approximating 
its direction as that of the soft pion, and parameterizing its momentum as 
a linear function of the soft-pion momentum, with parameters obtained 
from a Monte Carlo simulation. 
The presence of an undetected neutrino is inferred from conservation
of momentum and energy. The neutrino invariant mass squared is calculated as
\begin{eqnarray}
\Mnu \equiv (E_{\mbox{\rm beam}}-E_{{D^*}} - 
E_{\ell})^2-({\bf{p}}_{{D^*}} + {\bf{p}}_{\ell})^2\,,
\label{eqn:mms}
\end{eqnarray}
where $E_{\mbox{\rm beam}}$ is half the center-of-mass energy and $E_{\ell}~(E_{{D^*}})$ 
and ${\bf{p}}_{\ell}~({\bf{p}}_{{D^*}})$ are the center-of-mass energy and momentum 
of the lepton (the $D^*$ meson).  We set ${\bf{p}}_{B}$ = 0, which introduces a negligible
spread in $\Mnu$ compared with the approximation of the $D^{*}$ momentum based on the soft pion.
For signal decays that are properly reconstructed, the $\Mnu$ distribution peaks
near zero.  Background events, however, are spread over a wide range of $\Mnu$ values.  
We define a signal region ($\Mnu > -2~$GeV$^{2}/c^4$) and a sideband 
region ($-8 < \Mnu < -4~$GeV$^{2}/c^4$).

We use the symbol $\Ms$ to denote $\Mnu$ for any
candidate in the single-tag sample.
In the double-tag sample, we randomly choose one of the two
reconstructed $\Bzb \rightarrow D^{*+} \ell^{-}\bar{\nu}_{\ell}$ candidates 
as ``first'' and the other as ``second''. Their $\Mnu$ values are 
labeled $\M$ and $\MM$, respectively. We require that $\M$ fall in the signal region.

The single-tag and double-tag samples have several types of background:
continuum, combinatorial \BB, and peaking \BB.
The combinatorial \BB background originates from random combinations of
reconstructed leptons and soft pions.  
The peaking \BB background is composed of $\Bb \to D^{*} \pi \ell \bar{\nu}_{\ell}$ 
decays with or without an excited charmed resonance $D^{**}$~\cite{e961}, where the reconstructed 
soft pion comes from the decay $D^{*+}\to \Dz \pi^+$, leading to an
accumulation of these events at high values of $\Mnu$.
The peaking \BB background is suppressed by the requirement $p_\ell > 1.5\gevc$ 
on the lepton center-of-mass momentum.
Such events have an $\Mnu$ distribution that is different from the signal, allowing us 
to extract their contribution in the signal region.

The double-tag sample contains two additional types of background:
events in which the first candidate is combinatorial background and the second is signal
(called \M-combinatorial background) and events in which the first candidate is peaking
background and the second is signal (called \M-peaking background).

To determine \Ns\ and \Nd, we perform binned $\chi^2$ fits to 
one-dimensional histograms of the \Ms\ and \MM\ distributions of
on-resonance data events, ranging from $-8$ to $2~$GeV$^{2}/c^4$.
Before fitting, we subtract the continuum background contribution from
the histograms. This is determined using the \Ms\ and \MM\ distributions of
off-resonance data, scaled to account for the ratio of on-resonance to
off-resonance luminosities and the center-of-mass energy dependence of the
continuum production cross-section.
In addition, the contributions of the \M-combinatorial ($3\%$) and \M-peaking ($1\%$)
backgrounds are subtracted from the \MM\ histogram before doing the fit. 
The contribution of the \M-combinatorial background is determined from
sideband data.  The \M-peaking background is determined 
with simulated events.

After the subtraction, the \Ms\ and \MM\ histograms are fit
separately, to a function whose value for bin $j$ of the histogram is
\begin{equation}
f_j = \sum_t N^t P^t_j,
\end{equation}
where $N^t$ is the number of events of type $t$ ($t=$ signal,
combinatorial, peaking) populating the histogram, and $P^t_j$ is the
bin $j$ value of a discrete probability density function (PDF) obtained from
simulated events of type $t$, normalized such that 
$\sum_j P^t_j = 1$. 
The fit determines the parameters $N^t$ by minimizing
\begin{equation}
\chi^2 = \sum_j{{(H_j - f_j)^2 \over \sigma_{H_j}^2 + \sigma_{f_j}^2}},
\label{eq:fitchi2}
\end{equation}
where $H_j$ is the number of entries in bin $j$ of the data histogram being
fit; $\sigma_{H_j}$ is the statistical error on $H_j$, including
uncertainties due to the background subtractions described above; and
$\sigma_{f_j}$ is the error on $f_j$, determined from the errors on
$P_j^t$, which are due to the finite size of the simulated sample.

\begin{figure}[!htb]
\begin{center}
\vspace*{-2.7cm}
\includegraphics[width=8.29cm]{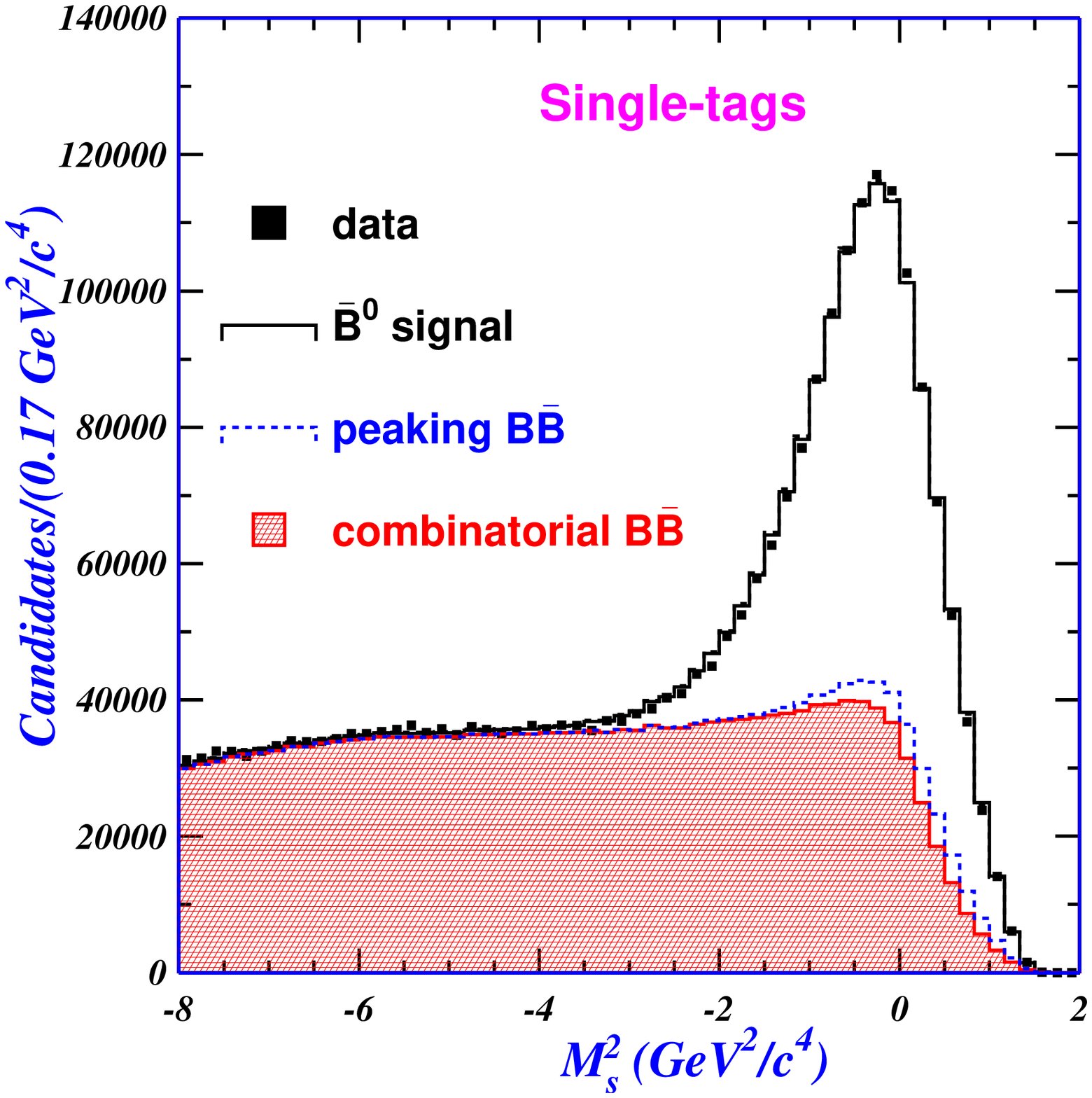}  
\vspace*{-3.3cm}
\hspace{-0.5cm}
\includegraphics[width=8.29cm]{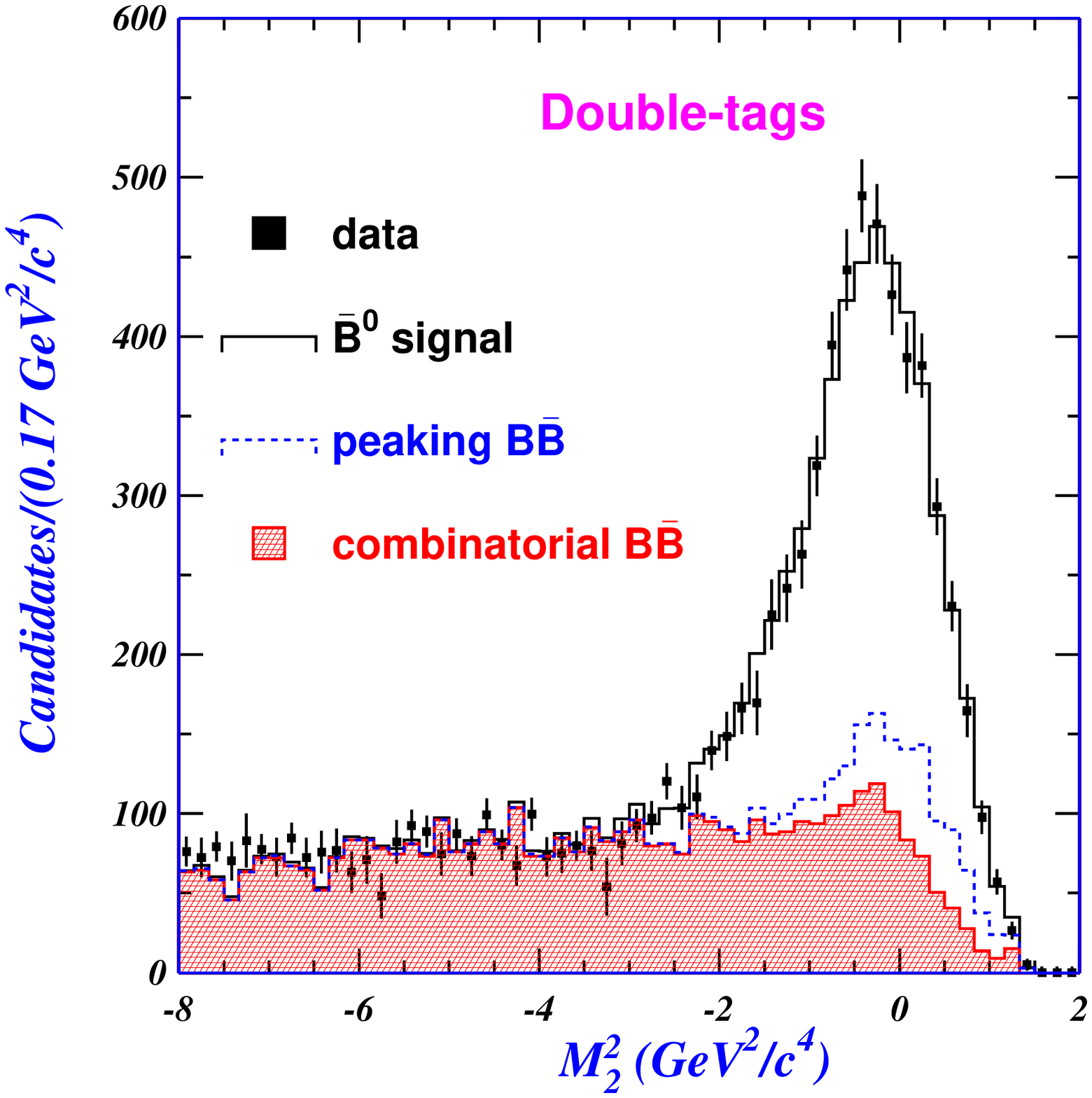}  
\vspace*{-0.8cm}
\caption{The \Ms\ (top) and \MM\ (bottom) distributions for the
on-resonance sample. The continuum background has been subtracted from
the distributions. In addition, the \M-combinatorial and the \M-peaking
backgrounds have been subtracted from the \MM\ distribution.
The levels of the simulated signal, peaking \BB, and combinatorial \BB 
background contributions are obtained from the fit.}
\label{fig:rightsign}
\end{center}
\end{figure}

The results of the fits are presented in Table~\ref{tab:numbers}.
The \Ms\ and \MM\ distributions are shown in
Fig.~\ref{fig:rightsign}.
The fits yield $\Ns = 786200 \pm 1900$ (Confidence Level (C.L.)= 11\%) 
and $\Nd = 3560 \pm 70$ (C.L. = 82\%).
Equation~(\ref{eq:f00}) then gives
$\fzz = 0.487 \pm 0.010$, where the error is due to data statistics.
\begin{table}[!htb]\tabcolsep=3mm
\caption{Numbers of entries of different types in the \Ms\ and \MM\ 
histograms in the signal region.}
\begin{center}      
\begin{tabular}{lcc} 
\hline \hline 
{\raisebox{-0.5ex}{Source}}  & {\raisebox{-0.5ex}{\Ms\ }}               
                             & {\raisebox{-0.5ex}{\MM\ }}     \\ [0.6mm]\hline 
Signal                 & $786200 \pm 1900$  & $3560 \pm 70$   \\ \hline
{\raisebox{-0.5ex}{Combinatorial $\BB$}}    
& {\raisebox{-0.5ex}{$558080 \pm 470$}}   
& {\raisebox{-0.5ex}{$1510 \pm 20$}}  \\ [0.3mm]
Peaking $\BB$          & $68170  \pm 260$   & $300  \pm 20$   \\
Continuum              & $240600 \pm 1400$  & $160  \pm 40$   \\
$\M$-combinatorial     & ---                & $180  \pm 20$   \\
$\M$-peaking           & ---                & $60   \pm 10$   \\ \hline \hline 
\end{tabular}
\end{center}
\label{tab:numbers}
\end{table}

\begin{figure}[!htb]
\begin{center}
\vspace*{-2.3cm}
\hspace*{-0.6cm}
\includegraphics[width=10.4cm]{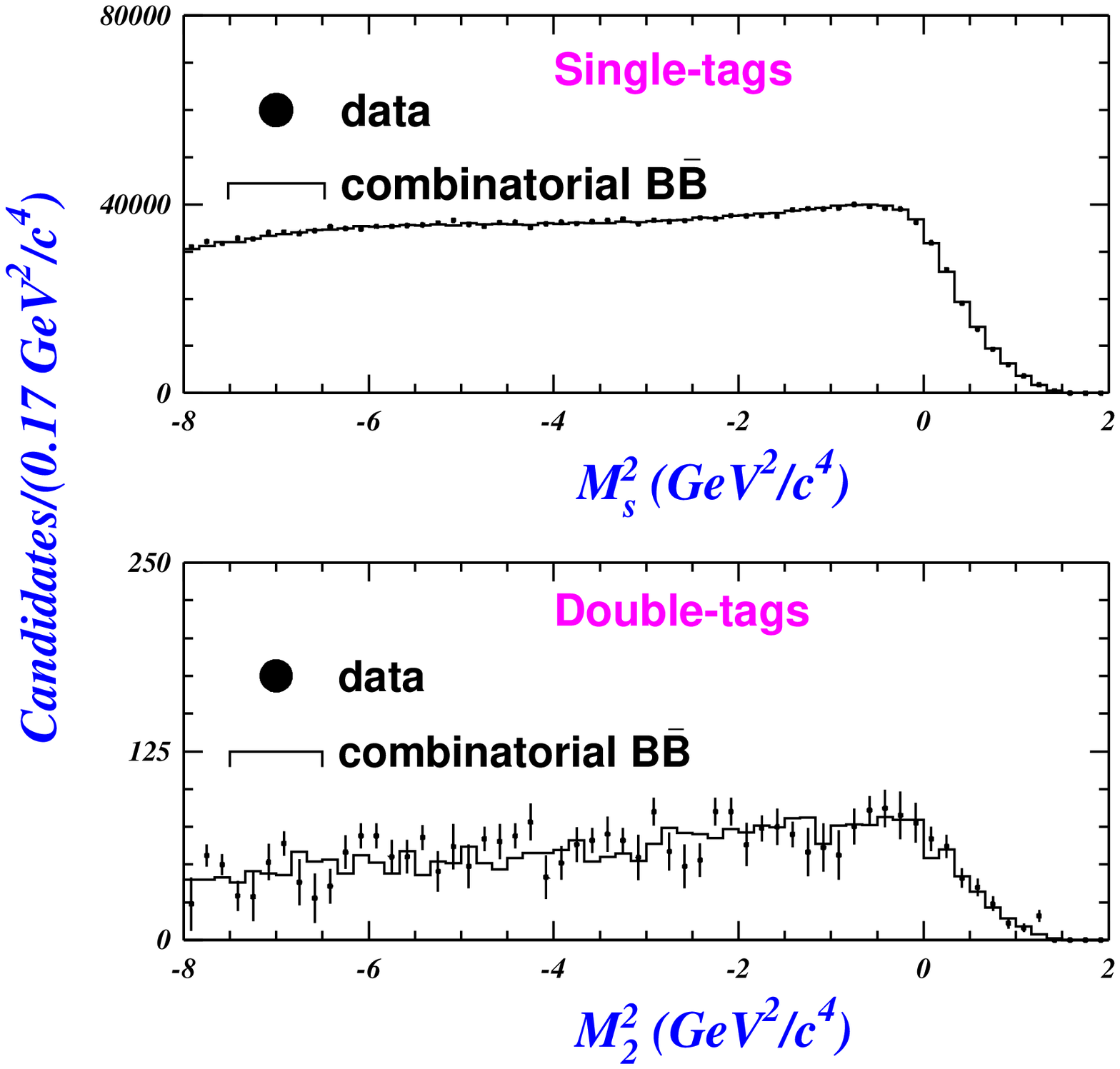}
\vspace*{-2.5cm}
\caption{The \Ms\ (top) and \MM\ (bottom) 
distributions for the same-charge on-resonance sample.
The continuum background has been subtracted from the distributions.
The \M-combinatorial and the \M-peaking backgrounds have been subtracted 
from the \MM\ distribution.  The level of the simulated combinatorial 
\BB\ background is obtained from the fit.}
\label{fig:wrongsign}
\end{center}
\end{figure}

To determine how well the simulation reproduces the \Ms\ and \MM\
distributions for the combinatorial background in the data,
we study the distributions for a sample of same-charge
candidates, in which the lepton and soft pion have the same
electric charge.
We fit the continuum-subtracted \Ms\ and \MM\ histograms of the
same-charge sample using the function $f'_j = N P'_j$, where $P'_j$ is
the bin $j$ value of the PDF of same-charge simulated \BB\ events, 
normalized such that $\sum_j P'_j =1$, and the parameter $N$ 
is determined by the fit.
The histograms, overlaid with the fit function, are shown in
Fig.~\ref{fig:wrongsign}.  The accumulated differences 
$D \equiv \sum_j (H'_j - f'_j)$
between the same-charge data histograms $H'_j$ and the fit
functions are summarized in Table~\ref{tab:ws_ratio}. 
Their consistency with zero indicates that the distributions of simulated
combinatorial \BB\ background events do not lead to significant fake
signal yields. Nevertheless, we evaluate a systematic uncertainty on 
the modeling of the combinatorial background based on the observed difference
in the like-sign sample.
\begin{table}[!htb]
\centering\caption{The difference $D \equiv \sum_j (H'_j - f'_j)$ between the
same-charge data histogram and the fit function, summed over the signal region
or over the whole region of the \Ms\ and \MM\ distributions.}
\begin{tabular}{lcccc}  
\hline \hline 
Fit           & \multicolumn{2}{c}{Signal region} 
              & \multicolumn{2}{c}{Whole region} \\
parameter     & \Ms     & \MM        &   \Ms        & \MM \\[0.3mm] \hline
$D$           & $-1300 \pm 2100$     & $-80 \pm 80$ \hspace{0.5mm}  
	      & $700   \pm 3000$     & $70  \pm 80$ \hspace{0.5mm}  \\
C.L.($\%$)    & 57      & 78         & 94           & 98  \\ \hline \hline
\end{tabular}
\label{tab:ws_ratio}
\end{table}

We evaluate the absolute systematic uncertainties in $f_{00}$ due to 
the \M-combinatorial subtraction (0.0005), the \M-peaking background (0.0005), 
the value of $C$ due to the track multiplicity dependence 
of the efficiency (0.0015), the finite size of the simulated sample (0.002),
the same-charge sample (0.0025), 
the impact of a possible contribution of non-$\BB$ decays of 
the $\Upsilon(4S)$~\cite{nonbbar} (0.0025), 
the peaking background composition (0.004), and 
the total number of $\BB$, \NBB\ (0.0055).

The dominant contribution to the systematic error comes from a $1.1\%$ systematic uncertainty 
in \NBB, due mainly to the uncertainty in the tracking efficiency.
The peaking \BB background is estimated from the simulated sample containing all $D^{**}$ 
resonances and non-resonant events.  We vary the ratio of the branching fraction of
the resonant and the non-resonant production such that the variation 
of this ratio is wide enough to include poorly known decays.  
We repeat the analysis procedure to determine \Ns\ and \Nd.  
The uncertainties due to the lepton and soft-pion momentum spectra are negligible.
We combine the uncertainties given above in quadrature to determine an
absolute systematic error of 0.008 for $\fzz$.

In summary, we use a partial reconstruction of the decay
$\Bzb \rightarrow D^{*+} \ell^{-}\bar{\nu}_{\ell}$ to obtain the result
\begin{equation}
\fzz = 0.487 \pm 0.010(stat) \pm 0.008(sys), 
\end{equation}
where the first error is statistical and the second is systematic.
This result is the first, precise, and direct measurement of \fzz. 
Since this measurement is made by comparing the numbers of events with
one and two reconstructed $\Bzb \rightarrow D^{*+} \ell^{-}\bar{\nu}_{\ell}$ decays, 
it does not depend on branching fractions of $\Bzb \rightarrow D^{*+} \ell^{-} \bar{\nu}_{\ell}$
and $D^{*+} \to \Dz \pi^+$ decays, on the ratio of the charged and neutral $B$ meson lifetimes, 
nor on the assumption of isospin symmetry.
By combining our $\fzz$ result with the world average of $R^{+/0}$ noted in the introduction, 
we add the errors quadratically to obtain $f_{+-} = 0.490 \pm 0.023$.  
Thus we find the fraction of $\Upsilon(4S) \rightarrow$ non-$\BB$ to be $1 - \fzz - f_{+-} = 0.023 \pm 0.032$.
If $\fzz + f_{+-} = 1$, our $\fzz$ result can be averaged with $R^{+/0}$~\cite{hfag05} 
to yield $\fzz = 0.494 \pm 0.008$, $f_{+-} = 0.506 \pm 0.008$, 
and $f_{+-} / \fzz = 1.023 \pm 0.032$. This value of $f_{+-} / \fzz$ is in good agreement 
with isospin conservation in $\Upsilon(4S) \rightarrow \BB$ within errors.

\begin{acknowledgments}
We are grateful for the excellent luminosity and machine conditions
provided by our \pep2\ colleagues, 
and for the substantial dedicated effort from
the computing organizations that support \babar.
The collaborating institutions wish to thank 
SLAC for its support and kind hospitality. 
This work is supported by
DOE
and NSF (USA),
NSERC (Canada),
IHEP (China),
CEA and
CNRS-IN2P3
(France),
BMBF and DFG
(Germany),
INFN (Italy),
FOM (The Netherlands),
NFR (Norway),
MIST (Russia), and
PPARC (United Kingdom). 
Individuals have received support from CONACyT (Mexico), A.~P.~Sloan Foundation, 
Research Corporation,
and Alexander von Humboldt Foundation.

\end{acknowledgments}

\end{document}